\documentclass[english]{article}
\usepackage[T1]{fontenc}
\usepackage[latin9]{inputenc}
\usepackage{geometry}
\usepackage{color}
\usepackage{amsmath}
\usepackage{amssymb}
\usepackage{graphicx}
\usepackage{wasysym}

\makeatletter

\providecommand{\tabularnewline}{\\}

\usepackage{jheppub}

\author[a,b]{Tamas Gombor}

\author[a]{Zoltán Bajnok}

\affiliation[a]{ HUN-REN Wigner Research Centre for Physics,\\ Konkoly-Thege Miklós u. 29-33, 1121 Budapest, Hungary}

\affiliation[b]{Department of Theoretical Physics, Eötvös Loránd University,\\ Pázmány Péter sétány 1/A, 1117 Budapest, Hungary}

\emailAdd{gombort@caesar.elte.hu}
\emailAdd{bajnok.zoltan@winger.hun-ren.hu}

\abstract{Integrable $su(2\vert2)_{c}$ symmetric models have integrable boundaries with $osp(2\vert2)$ symmetries, which can be embedded into $su(2\vert2)_{c}$ in two different ways.
We dualize the previously obtained asymptotic overlap formulas for one of the embeddings to describe the other embedding and apply the results to describe the asymptotic expectation values of local operators in the presence of a 't Hooft line in ${\cal N}=4$ SYM.
A peculiar feature of the setting is that in certain gradings only descendant states have non-vanishing overlaps with the boundary state and the overlap formula is not factorized for the Bethe roots.} 


\makeatother

\usepackage{babel}
\begin{document}
\title{Dual overlaps and finite coupling 't Hooft loops}
\maketitle

\section{Introduction}

Integrable boundaries have been getting considerable interest recently
in statistical physics as well as in the AdS/CFT correspondence. The
developments were motivated by quench problems in spin chains on the
statistical physics side \cite{Caux:2013ra,De_Nardis_2014,Piroli:2018ksf,Piroli:2018don,Rylands:2022gev},
while in the AdS/CFT correspondence they describe various conformal
defects in the gauge theory and related probe branes in the string
theory \cite{Buhl-Mortensen:2015gfd,Jiang:2019xdz,Jiang:2019zig,Yang:2021hrl,Kristjansen:2021abc,Jiang:2023cdm,Kristjansen:2023ysz,Ivanovskiy:2024vel}.
In most of the applications the physically relevant quantities are
the overlaps of multi-particle finite volume states with integrable
boundary states \cite{DeLeeuw:2018cal,deLeeuw:2019ebw,Gombor:2022aqj}.

The prototypical AdS/CFT duality relates the ${\cal N}=4$ supersymmetric
Yang-Mills (SYM) theory to strings propagating on the $AdS_{5}\times S^{5}$
background \cite{Beisert:2010jr}. The total superconformal symmetry
of the conformal field theory $psu(2,2\vert4)$ is manifested as isometries
of the supercoset string sigma model. By introducing a codimension
1 defect in the CFT, with prescribed boundary conditions for the fields,
one can preserve half of the supersymmetry \cite{Karch:2000gx}. The
defect breaks partially the translational symmetry, which implies
that single trace operators acquire non-trivial vacuum expectation
values (VEVs). The space-time dependence of the VEV is completely
determined in terms of the scaling dimension, but its proportionality
factor is a non-trivial function of the 't Hooft coupling. As single
trace operators corresponds to multi-particle finite volume states
the proportionality factor turns out to be the overlap of this state
with an integrable boundary state \cite{Buhl-Mortensen:2015gfd,deLeeuw:2015hxa}.

Integrable boundary states have to satisfy severe consistency requirements
including the boundary Yang-Baxter equations for their K-matrix \cite{Piroli:2017sei,Gombor:2020kgu,Gombor:2021hmj}.
These equations involve the scattering matrix of the excitations,
which has a factorized form with $su(2\vert2)_{c}\oplus su(2\vert2)_{c}$
symmetry, the remnant of $psu(2,2\vert4)$ after fixing the gauge
and quantizing string theory. In our works \cite{Gombor:2020kgu,Gombor:2020auk}
we classified the solutions for integrable factorized K-matrices (one
of these solutions were also found in \cite{Komatsu:2020sup} independently),
which came in two versions, both having $osp(2\vert2)$ symmetry,
such that together with $su(2\vert2)_{c}$ they form a symmetric pair.
The difference lies in the embedding of this symmetry into $su(2\vert2)_{c}$
, i.e. whether the bosonic $su(2)$ part of $osp(2\vert2)$ lies in
the Lorentzian or in the R-symmetry part. The codimension 1 defect
implies that it is in the Lorentzian part, thus in order to describe
the expectation values of single trace operators we developed a procedure
and calculated the corresponding overlaps for this type of boundary
states. The result is very general and applies in all the cases with
the same symmetry.

Recently a 't Hooft line was introduced in the CFT, i.e. a codimension
3 defect and its integrability properties were investigated \cite{Kristjansen:2023ysz}.
Its presence implies specific boundary conditions for the SYM fields
and the expectation values of single trace operators were connected
to overlaps of multiparticle states with a boundary state, which seemed
to be integrable. This was supported by directly calculating the overlaps
in various subsectors at leading order. Since the geometrical setting
breaks the symmetries as expected for the other $osp(2\vert2)$ embedding
we expect that the all loop overlaps should be calculated from our
integrable K-matrix. The aim of our paper is to calculate the asymptotic
overlaps for the corresponding K-matrix.

Since the two K-matrices are related by a duality transformation we
need to understand how duality acts on $su(2\vert2)_{c}$ overlaps.
It turns out that due to the specific selection rules, highest weight
Bethe states have vanishing overlaps with the boundary state and only
their descendants can have non-zero overlaps. In order to investigate
this phenomena we analyze the rational version of the problem, i.e.
dualities for overlaps in rational $su(2\vert2)$ spin chains and
their action on overlaps. This is also useful to make contact with
the LO 't Hooft loop calculations.

The paper is organized as follows: First, in Section 2 we recall the
integrable K-matrices for an $su(2\vert2)_{c}$ symmetric scattering
matrix. Having introduced the Bethe ansatz equations and integrable
K-matrices we present the previously calculated asymptotic overlaps
for one of these K-matrices. We then calculate its weak coupling,
rational limit. In Section 3 we analyze the two types of rational
K-matrices and the overlaps in various gradings. We pay particular
attention to degenerate cases. In Section 4 we elevate the results
for $gl(4\vert4)$. The various fermionic dualities enables us to
make correspondences with all the existing overlaps in the 't Hooft
loop setting. In Section 5, using a fermionic duality, we determine
the overlap formulas for the other type of $K$-matrix. These results
are then used in Section 6 to describe the asymptotic overlaps for
the 't Hooft loop. We conclude in Section 7. We added several Appendices,
in which we investigate the condition whether the Bethe state can
have a non-vanishing overlap with the boundary state, and how to calculate
the overlaps of descendent states in the various cases.

\section{K-matrices, overlaps and their weak coupling limit}

In the prototypical duality there are 8 fermionic and 8 bosonic excitations,
which scatters on each other in an integrable way. Due to the factorized
$su(2\vert2)_{c}\oplus su(2\vert2)_{c}$ symmetry the scattering matrix
has also a factorized form 
\begin{equation}
\mathbb{S}(p_{1},p_{2})=S_{0}(p_{1},p_{2})S(p_{1},p_{2})\otimes S(p_{1},p_{2})\label{eq:Smatrix}
\end{equation}
The centrally extended $su(2|2)_{c}$ symmetry completely fixes the
matrix structure of the scattering matrix $S(p_{1},p_{2})$ \cite{Beisert:2005tm}
of the particles labeled by $(1,2\vert3,4)$, where $1,2$ are considered
to be bosonic, while $3,4$ as fermionic. Their dispersion relation
is also fixed by the symmetry
\begin{equation}
\epsilon(p_{j})=ig\biggl(x_{j}^{-}-\frac{1}{x_{j}^{-}}-x_{j}^{+}+\frac{1}{x_{j}^{+}}\biggr)\quad;\quad x_{j}^{\pm}=e^{\pm i\frac{p_{j}}{2}}\frac{1+\sqrt{1+16g^{2}\sin^{2}\frac{p_{j}}{2}}}{4g\sin\frac{p_{j}}{2}}
\end{equation}
where the 't Hooft coupling is $\lambda=16\pi^{2}g$ and the momentum
of the particles is simply $e^{ip_{j}}=\frac{x_{j}^{+}}{x_{j}^{-}}.$

\subsection{Periodic $su(2\vert2)_{c}$ spin chains}

In calculating the asymptotic, large volume ( R-charge) spectrum one
has to diagonalize an inhomogeneous $su(2|2)_{c}$ transfer matrix
\begin{equation}
t(p_{0})={\rm Tr}_{0}(S(p_{0},p_{1})\dots S(p_{0},p_{2L}))
\end{equation}

This can be done by the Bethe ansatz method (which could be coordinate
or algebraic). As a first step one finds a simple eigenstate of the
transfer matrix: all state being the same, which can be either bosonic
or fermionic. Typically we choose the 1s, and denote this bosonic
pseudovacuum as $\vert0^{b}\rangle=\otimes^{2L}\vert1\rangle$. One
can then introduce excitations, by turning 1s into other labels and
build a plane wave out of them. The $y$ roots are the momenta of
the plane waves of excitations of $3$s. Finally, $w$ labels the
momenta when $4$s are created by flipping $3$s. The eigenvalue is
then labeled by the set of magnons: $y$-roots $\{y_{k}\}_{k=1..N}$
and $w$-roots $\{w\}_{l=1..M}$ as 
\begin{equation}
t(u)\vert\mathbf{y},\mathbf{w}\rangle=\Lambda(u,\mathbf{y},\mathbf{w})\vert\mathbf{y},\mathbf{w}\rangle
\end{equation}
where we reparametrized $p_{0}$ with $u$ as $x_{0}^{\pm}+\frac{1}{x_{0}^{\pm}}=u\pm\frac{i}{2g}$.
The eigenvalue $\Lambda(u,\mathbf{y},\mathbf{w})$ can be written
as 
\begin{equation}
\Lambda(u,\mathbf{y},\mathbf{w})=e^{i\frac{p_{0}(u)}{2}(N-2L)}\frac{\mathcal{R}^{(+)+}}{\mathcal{R}^{(+)-}}\left\{ \frac{\mathcal{R}^{(-)+}\mathcal{R}_{y}^{-}}{\mathcal{R}^{(+)+}\mathcal{R}_{y}^{+}}-\frac{\mathcal{R}_{y}^{-}Q_{w}^{++}}{\mathcal{R}_{y}^{+}Q_{w}}-\frac{\mathcal{B}_{y}^{+}Q_{w}^{--}}{\mathcal{B}_{y}^{-}Q_{w}}+\frac{\mathcal{B}^{(+)-}\mathcal{B}_{y}^{+}}{\mathcal{B}^{(-)-}\mathcal{B}_{y}^{-}}\right\} 
\end{equation}
where 
\begin{equation}
\mathcal{R}_{y}(u)=\prod_{j=1}^{N}(x(u)-y_{j})\quad;\quad;\quad Q_{w}(u)=\prod_{l=1}^{M}(u-w_{l})\quad;\qquad\mathcal{R}^{(\pm)}(u)=\prod_{j=1}^{2L}(x(u)-x_{j}^{\pm}).
\end{equation}
and the $\mathcal{B}$ quantities can be obtained from the $\mathcal{R}$-s
by replacing $x(u)$ with $1/x(u)$: 
\begin{equation}
\mathcal{B}_{y}(u)=\prod_{j=1}^{N}(1/x(u)-y_{j})\quad;\qquad\mathcal{B}^{(\pm)}(u)=\prod_{j=1}^{2L}(1/x(u)-x_{j}^{\pm})
\end{equation}
Shifts are understood as $f^{\pm}(u)=f(u\pm\frac{i}{2g})$ and we
assumed that the total momentum vanishes: $\sum_{j}p_{j}=0$, which
is enforced by the level matching condition. Finally, $x(u)$ is defined
by $x(u)+1/x(u)=u$.

The actual values for the roots can be obtained from the Bethe ansatz
equations, which originate from the regularity of the transfer matrix.
Indeed, the transfer matrix has to be regular at $x^{+}(u)=y_{j}$,
which implies the quantization conditions 
\begin{equation}
\frac{\mathcal{R}_{y}^{-}}{\mathcal{R}_{y}^{+}}\left\{ \frac{\mathcal{R}^{(-)+}}{\mathcal{R}^{(+)+}}-\frac{Q_{w}^{++}}{Q_{w}}\right\} _{x^{+}(u)=y_{j}}=\mathrm{regular}\quad\longrightarrow\quad\left\{ \frac{\mathcal{R}^{(-)}}{\mathcal{R}^{(+)}}-\frac{Q_{w}^{+}}{Q_{w}^{-}}\right\} _{x(u)=y_{j}}=0
\end{equation}
The corresponding Bethe ansatz equation reads explicitly as 
\begin{equation}
e^{i\phi_{v_{j}}}:=\prod_{k=1}^{2L}e^{-ip_{k}/2}\frac{y_{j}-x_{k}^{+}}{y_{j}-x_{k}^{-}}\prod_{k=1}^{M}\frac{v_{j}-w_{k}+\frac{i}{2g}}{v_{j}-w_{k}-\frac{i}{2g}}=1
\end{equation}
where $v_{j}=y_{j}+y_{j}^{-1}$. Similarly, the $w$-roots are quantized
via the regularity at $u=w_{l}$ 
\begin{equation}
\left\{ \frac{\mathcal{R}_{y}^{-}Q_{w}^{++}}{\mathcal{R}_{y}^{+}}+\frac{\mathcal{B}_{y}^{+}Q_{w}^{--}}{\mathcal{B}_{y}^{-}}\right\} _{u=w_{l}}=0
\end{equation}
Using that 
\begin{equation}
{\cal Q}_{v}:=\frac{{\cal R}_{y}{\cal B}_{y}}{\prod_{j}(-y_{j})}=\prod_{j=1}^{N}(u-v_{j})
\end{equation}
we can write the corresponding BA equation as 
\begin{equation}
e^{i\phi_{w_{j}}}:=\prod_{k=1}^{N}\frac{w_{j}-v_{k}+\frac{i}{2g}}{w_{j}-v_{k}-\frac{i}{2g}}\prod_{\begin{array}{c}
l=1\\
l\neq j
\end{array}}^{M}\frac{w_{j}-w_{l}-\frac{i}{g}}{w_{j}-w_{l}+\frac{i}{g}}=1
\end{equation}

We note that the Bethe ansatz method provides eigenstates, which are
highest weight states for the symmetry algebra. Descendent states
can be obtained by applying symmetry transformations. They show up
in the Bethe ansatz equations as roots at infinities. The Bethe ansatz
equations assumed a specific order of creating excitations ($1\to3,3\to4$),
i.e. they are valid in a specific \emph{grading}. Later we explain
how to switch between different gradings.

\subsection{Integrable boundary states and overlaps}

Integrable boundaries can be represented as boundary states in spin
chains \cite{Piroli:2017sei}. They can be parametrized by K-matrices,
which satisfy the boundary K-Yang Baxter equations involving the scattering
matrix $S$. These equations are very restrictive and in the $su(2\vert2)_{c}$
case we found two types of solutions \cite{Gombor:2020kgu}:
\begin{equation}
K^{(1)}=\left(\begin{array}{cccc}
k_{1} & k_{2}+e^{(1)} & 0 & 0\\
k_{2}-e^{(1)} & k_{4} & 0 & 0\\
0 & 0 & 0 & f^{(1)}\\
0 & 0 & -f^{(1)} & 0
\end{array}\right),\quad K^{(2)}=\left(\begin{array}{cccc}
0 & f^{(2)} & 0 & 0\\
-f^{(2)} & 0 & 0 & 0\\
0 & 0 & k_{1} & k_{2}+e^{(2)}\\
0 & 0 & k_{2}-e^{(2)} & k_{4}
\end{array}\right).\label{eq:K12}
\end{equation}
where
\begin{equation}
e^{(1)}(p)=\frac{-i}{x_{s}}\frac{x^{+}+x_{s}^{2}x^{-}}{1+x^{+}x^{-}},\quad f^{(1)}(p)=\frac{i}{x_{s}}\sqrt{\frac{x^{-}}{x^{+}}}\frac{(x^{+})^{2}-x_{s}^{2}}{1+x^{+}x^{-}},
\end{equation}
with $k_{1}k_{4}-k_{2}^{2}=1,$ and
\begin{equation}
e^{(2)}(p)=\frac{i}{x_{s}}\frac{x^{-}+x_{s}^{2}x^{+}}{1+x^{+}x^{-}},\quad f^{(2)}(p)=\frac{i}{x_{s}}\sqrt{\frac{x^{+}}{x^{-}}}\frac{(x^{-})^{2}-x_{s}^{2}}{1+x^{+}x^{-}}.
\end{equation}
and some parameters $k_{i}$ and $x_{s}$. Both matrices have $osp(2\vert2)$
symmetry, but they differ in which way this symmetry is embedded into
$su(2\vert2)_{c}$.

Every solution of the K-Yang Baxter equation leads to an integrable
boundary state. We first create a two-site state and then distribute
it homogeneously aloung the chain
\begin{equation}
\langle\Psi_{K}\vert=\langle\psi_{1}\vert\otimes\langle\psi_{2}\vert\otimes\dots\otimes\langle\psi_{L}\vert\quad;\qquad\langle\psi\vert=\sum_{a,b}\langle a\vert\otimes\langle b\vert K_{ab}
\end{equation}

We investigated the overlap corresponding to the boundary state built
from $K^{(1)}$ in detail and found that the non-vanishing overlap
requires the following selection rules for the roots. The momenta
of the particles, i.e. the inhomogeneities of the spin chain, should
come in pairs $p_{i}=-p_{2L+1-i}$ as well as the $v$-s, $v_{i}=-v_{N+1-i}$
and $w$-s, $w_{i}=-w_{M+1-i}$, and their numbers should be related
as $N=2M$. In this case the overlap takes the form \cite{Gombor:2020kgu}
\begin{equation}
\frac{|\langle\Psi_{K^{(1)}}|\mathbf{y},\mathbf{w}\rangle|^{2}}{\langle\mathbf{y},\mathbf{w}|\mathbf{y},\mathbf{w}\rangle}=k_{1}^{2L-N}x_{s}^{-N}\frac{\mathcal{R}_{y}(x_{s})^{2}}{\mathcal{R}_{y}(0)}\frac{1}{Q_{w}(0)Q_{w}(\frac{i}{2g})}\frac{\det G^{+}}{\det G^{-}},
\end{equation}
where $G^{\pm}$ are related to the factorization of the Gaudin determinant
for the paired state 
\begin{equation}
G=\left|\begin{array}{cc}
\left(\partial_{v_{i}}\phi_{v_{j}}\right)_{N\times N} & \left(\partial_{v_{i}}\phi_{w_{j}}\right)_{N\times M}\\
\left(\partial_{w_{i}}\phi_{v_{j}}\right)_{M\times N} & \left(\partial_{w_{i}}\phi_{w_{j}}\right)_{M\times M}
\end{array}\right|=G_{+}G_{-}
\end{equation}
These formulae are valid for even $M$. For odd $M$ the function
$Q_{w}(u)$ has a zero root $w_{\frac{M+1}{2}}=0$, $Q_{w}(u)=u\prod_{j=1}^{\frac{M-1}{2}}(u^{2}-w_{j}^{2})=u\bar{Q}_{w}(u)$.
In this case $\bar{Q}_{w}$ has to be used, instead of $Q_{w}$. Also
the factorization of the Gaudin determinant involves this special
root in a particular way, see \cite{Brockmann_2014odd,Gombor:2020kgu,Gombor:2023bez}
for details.

Our result is very general and describes overlaps in various physical
situations. The different cases are distinguished by how $x_{s}$
depends on the coupling constant.

\subsection{Weak coupling limit}

At weak coupling the $su(2\vert2)_{c}$ symmetry reduces to $su(2\vert2)$
and the inhomogeneous spin chain becomes rational. In this spin chain
we have three different type of roots $u^{(i)}$, which describe how
we flip the labels starting from all the $1$s, and three different
type of $Q_{i}$ functions, $Q_{i}=\prod_{j=1}^{N_{i}}(u-u_{j}^{(i)})$.
The $y$ type roots can scale at weak coupling either as $y\sim\frac{u^{(1)}}{g}$
or as $y\sim\frac{g}{u^{(3)}}$ distinguishing between type $1$ and
type $3$ roots. The $w$ roots scale as $w\sim\frac{u^{(2)}}{g}$
and they become the type $2$ roots. A type 1 root $u^{(1)}$ describes
how to create 3s from 1s, $u^{(2)}$ creates 4s from 3s, while $u^{(3)}$
creates 2s from 4s. They correspond to a specific grading.

The limit of the K-matrix depends on the behavior of $x_{s}$ at small
coupling. If $x_{s}$ does not depend on $g$, the $K$-matrix in
the weak coupling limit reads as
\begin{equation}
K^{(1)}(p)=\left(\begin{array}{cccc}
k_{1} & k_{2} & 0 & 0\\
k_{2} & k_{4} & 0 & 0\\
0 & 0 & 0 & \frac{i}{x_{s}}\\
0 & 0 & -\frac{i}{x_{s}} & 0
\end{array}\right),
\end{equation}
and the limit of the overlap formula is
\begin{equation}
\frac{|\langle\Psi_{K^{(1)}}|\mathbf{u}\rangle|^{2}}{\langle\mathbf{u}|\mathbf{u}\rangle}=k_{1}^{2L-N_{1}-N_{3}}x_{s}^{N_{3}-N_{1}}\frac{Q_{1}(0)Q_{3}(0)}{Q_{2}(0)Q_{2}(i/2)}\frac{\det G^{+}}{\det G^{-}}.
\end{equation}
If, however, $x_{s}$ is defined in a $g$-dependent way, say as $x_{s}+x_{s}^{-1}=\frac{is}{g}$,
then the $K$-matrix in the weak coupling limit is degenerate
\begin{equation}
K^{(1)}=\left(\begin{array}{cccc}
k_{1} & k_{2}+\frac{s}{u+i/2} & 0 & 0\\
k_{2}-\frac{s}{u+i/2} & k_{4} & 0 & 0\\
0 & 0 & 0 & 0\\
0 & 0 & 0 & 0
\end{array}\right).
\end{equation}
and the limit of the overlap formula is
\begin{equation}
\frac{|\langle\Psi_{K^{(1)}}|\mathbf{u}\rangle|^{2}}{\langle\mathbf{u}|\mathbf{u}\rangle}=k_{1}^{2L-2N_{1}}\frac{Q_{1}(is)^{2}Q_{3}(0)}{Q_{1}(0)Q_{2}(0)Q_{2}(i/2)}\frac{\det G^{+}}{\det G^{-}},
\end{equation}
with the selection rules $N_{1}=N_{2}=N_{3}$. These results are valid
in a specific grading.

We can easily calculate the weak coupling limit of the second type
of boundary K-matrix, $K^{(2)}$ and realize that $K^{(1)}$ and $K^{(2)}$
are related by the $1\leftrightarrow3,2\leftrightarrow4$ flips. This,
however, is nothing but choosing a different grading. In the following
we investigate how changing the grading will transform the overlap
formulas in order to describe the overlaps with the K-matrix $K^{(2)}$.

\section{Overlaps and dualities for rational $su(2|2)$ spin chains}

First we recall the overlaps in the $su(2)$ spin chains as we will
encounter the same quantities later when we analyze the $su(2\vert2)$
spin chain.

The generic $\mathfrak{su}(2)$ K-matrix is nothing but the one, which
appeared in the weak coupling limit in a 2 by 2 box and has the form
\begin{equation}
K(u)=\left(\begin{array}{cc}
k_{1} & k_{2}+\frac{s}{u+i/2}\\
k_{2}-\frac{s}{u+i/2} & k_{4}
\end{array}\right)
\end{equation}
The corresponding boundary state has the following overlap with the
Bethe states \cite{Gombor:2021uxz,Pozsgay_2018}
\begin{equation}
\frac{\vert\langle\Psi_{K}|\mathbf{u}\rangle\vert^{2}}{\langle\mathbf{u}|\mathbf{u}\rangle}\sim\frac{Q_{1}(is)^{2}}{Q_{1}(0)Q_{1}(\frac{i}{2})}\frac{\det G^{+}}{\det G^{-}}
\end{equation}
In this section we do not pay attention on scalar prefactors. The
overlap has two interesting limits. For $s\to0$ the K-matrix becomes
symmetric and the overlap takes the form 
\begin{equation}
K(u)=\left(\begin{array}{cc}
k_{1} & k_{2}\\
k_{2} & k_{4}
\end{array}\right)\quad;\quad\frac{\vert\langle\Psi_{K}|\mathbf{u}\rangle\vert^{2}}{\langle\mathbf{u}|\mathbf{u}\rangle}\sim\frac{Q_{1}(0)}{Q_{1}(\frac{i}{2})}\frac{\det G^{+}}{\det G^{-}}
\end{equation}
while in the opposite $s\to\infty$ limit the K-matrix is anti-symmetric
with the overlap
\begin{equation}
K(u)=\left(\begin{array}{cc}
0 & 1\\
-1 & 0
\end{array}\right)\quad;\quad\frac{\vert\langle\Psi_{K}|\mathbf{u}\rangle\vert^{2}}{\langle\mathbf{u}|\mathbf{u}\rangle}\sim\frac{1}{Q_{1}(0)Q_{1}(\frac{i}{2})}\frac{\det G^{+}}{\det G^{-}}
\end{equation}
Let us turn now to the $su(2|2)$ case.

\subsection{Gradings, nesting and overlaps in the regular case}

For the rational $su(2|2)$ spin chains there exist two types of integrable
$K$-matrices corresponding to the two different realizations of the
unbroken $osp(2\vert2)$ symmetry and they come with chiral pair structures:
\begin{equation}
K^{(1)}=\left(\begin{array}{cccc}
k_{1} & k_{2} & 0 & 0\\
k_{2} & k_{4} & 0 & 0\\
0 & 0 & 0 & -s\\
0 & 0 & s & 0
\end{array}\right),\qquad K^{(2)}=\left(\begin{array}{cccc}
0 & -s & 0 & 0\\
s & 0 & 0 & 0\\
0 & 0 & k_{1} & k_{2}\\
0 & 0 & k_{2} & k_{4}
\end{array}\right).\label{eq:ratK}
\end{equation}
They are related to the $g\to0$ limits of the two K-matrices for
the $su(2\vert2)_{c}$ case. We fix the normalization of the K-matrices
by demanding that $k_{1}k_{4}-k_{2}^{2}=1$. These two K-matrices
are related to each other by the $1\leftrightarrow3,2\leftrightarrow4$
changes. As these changes are also related to nestings and gradings
we recall them now.

In the diagonalization of the super spin chain transfer matrices we
can use different paths for the nesting. Let us choose the pseudo
vacuum as $A_{1}$ and the magnons $u^{(1)},u^{(2)},u^{(3)}$ such
a way, that they change the flavors as $A_{1}\to A_{2},A_{2}\to A_{3},A_{3}\to A_{4}$,
respectively, where $A_{k}\in\{1,2,3,4\}$ all distinct. We denote
this nesting as $(A_{1},A_{2},A_{3},A_{4})$. In particular, for the
nesting $(1,2,3,4)$ the Bethe state with magnon numbers $N_{1},N_{2},N_{3}$
will have the following number of $1,2,3$ and $4$ labels: $\#1=L-N_{1}$,
$\#2=N_{1}-N_{2}$, $\#3=N_{3}-N_{4}$ and $\#4=N_{3}$. This nesting
corresponds to the grading $(++--)$ and the Dynkin-diagram $\Circle-\otimes-\Circle$,
where $\pm$ corresponds to bosonic/ferminic indices and the bosonic
$\Circle$ versus fermionic $\otimes$ nodes encode if there is a
change in the nature of the labels. Another example is the nesting
$(1,3,4,2)$ where $\#1=L-N_{1}$, $\#2=N_{3}$, $\#3=N_{1}-N_{2}$
and $\#4=N_{2}-N_{3}$. The grading is $(+--+)$, while the corresponding
Dynkin-diagram is $\otimes-\Circle-\otimes$. Bethe states depend
on the nesting.

The overlap of a Bethe state with a boundary state depends both on
the nesting and the type of the K-matrix. For the nesting $(A_{1},A_{2},A_{3},A_{4})$
and K-matrix $K^{(i)}$ we denote it by
\begin{equation}
S_{i}^{(A_{1},A_{2},A_{3},A_{4})}(\mathbf{u})=\frac{|\langle\Psi_{i}|\mathbf{u}\rangle|^{2}}{\langle\mathbf{u}|\mathbf{u}\rangle}
\end{equation}
Since the two $K$-matrices are connected by the changes $1\leftrightarrow3$;
$2\leftrightarrow4$, the overlaps are also related 
\begin{equation}
S_{2}^{(A_{1},A_{2},A_{3},A_{4})}(\mathbf{u})=S_{1}^{(\bar{A}_{1},\bar{A}_{2},\bar{A}_{3},\bar{A}_{4})}(\mathbf{u}),\label{eq:idOV}
\end{equation}
where $\bar{1}=3;\bar{2}=4;\bar{3}=1;\bar{4}=2$ . The above changes
leave invariant the S-matrix implying that the Bethe equations and
the Bethe roots are the same on the two sides. We have already obtained
the overlap $S_{1}^{(1,3,4,2)}$ from the $g\to0$ limit as 
\begin{equation}
S_{1}^{(1,3,4,2)}\sim\frac{Q_{1}(0)Q_{3}(0)}{Q_{2}(0)Q_{2}(i/2)}\frac{\det G^{+}}{\det G^{-}},\label{eq:alap}
\end{equation}
with the selection rule $N_{1}+N_{3}=2N_{2}$. By using the relation
(\ref{eq:idOV}) we can easily write the overlap for the other type
of $K$-matrix in the opposite grading
\begin{equation}
S_{2}^{(3,1,2,4)}\sim\frac{Q_{1}(0)Q_{3}(0)}{Q_{2}(0)Q_{2}(i/2)}\frac{\det G^{+}}{\det G^{-}}.\label{eq:start}
\end{equation}

The natural question is, how we could describe this second overlap
in the original grading or in any other gradings. In order to get
this result, we have to apply fermionic dualities \cite{Kristjansen:2020vbe}.
The fermionic duality expresses the Bethe roots $\mathbf{u}$ in a
given grading with the dual Bethe roots $\tilde{\mathbf{u}}$ in the
dual grading. The original overlap written in terms of the dual roots
provides the overlap in the dual grading $S_{2}^{\mathbf{A}}(\mathbf{u})=S_{2}^{\tilde{\mathbf{A}}}(\tilde{\mathbf{u}})$.
There is, however an important difference. Although the original overlap
corresponds to a (highest weight) Bethe state $\vert\mathbf{u}\rangle$
the dual overlap corresponds only to a descendent state, which can
obtained by acting with some fermionic generator $|\mathbf{u}\rangle=\mathbb{Q}|\tilde{\mathbf{u}}\rangle$
on the h.w. Bethe state $|\tilde{\mathbf{u}}\rangle$ in the dual
grading. This implies that 
\begin{equation}
\frac{|\langle\Psi_{K^{(2)}}|\mathbf{u}\rangle|^{2}}{\langle\mathbf{u}|\mathbf{u}\rangle}=S_{2}^{\mathbf{A}}(\mathbf{u})=S_{2}^{\tilde{\mathbf{A}}}(\tilde{\mathbf{u}})=\frac{|\langle\Psi_{K^{(2)}}|\mathbb{Q}|\tilde{\mathbf{u}}\rangle|^{2}}{|\mathbb{Q}|\tilde{\mathbf{u}}\rangle|^{2}}\label{eq:overlapduality}
\end{equation}
 In order to calculate this (descendant) overlap we use the fermionic
dualities. The duality transformation involves two steps: the changing
of the Gaudin determinant and using the $QQ$-relations. The Gaudin
determinant transforms as
\begin{equation}
\frac{\det G^{+}}{\det G^{-}}\sim\begin{cases}
\frac{Q_{a-1}(i/2)Q_{a+1}(i/2)}{\bar{Q}_{a}(0)\tilde{\bar{Q}}_{a}(0)}\frac{\det\tilde{G}^{+}}{\det\tilde{G}^{-}}, & \text{if }N_{a}=\mathrm{even}\\
\frac{\bar{Q}_{a}(0)\tilde{\bar{Q}}_{a}(0)}{Q_{a-1}(i/2)Q_{a+1}(i/2)}\frac{\det\tilde{G}^{+}}{\det\tilde{G}^{-}}, & \text{if }N_{a}=\mathrm{odd}
\end{cases}
\end{equation}
where $\tilde{Q}_{a}$ denotes the $Q$-function after the duality
transformation, for the new type of $a$ root. The $QQ$-relation
reads as 
\begin{equation}
Q_{a}(u)\tilde{Q}_{a}(u)\sim Q_{a+1}(u+i/2)Q_{a-1}(u-i/2)-Q_{a+1}(u-i/2)Q_{a-1}(u+i/2)
\end{equation}
In showing these relations, one has to be careful and investigate
the role of zero roots. Their presence depends on the parity of $N_{a-1}+N_{a+1}$
and they imply non-trivial relations between $Q$-functions. For instance,
if $N_{a-1}+N_{a+1}$ is odd then $\bar{Q}_{a}(0)\tilde{\bar{Q}}_{a}(0)\sim Q_{a+1}(i/2)Q_{a-1}(i/2)$
and the Gaudin determinants are proportional to each other. A careful
analysis shows that both $N_{1}$ and $N_{3}$ should be even and
$N_{1}+N_{3}=2N_{2}$. In this case the following transformation rule
can be applied:

\begin{equation}
Q_{a}(0)\to\tilde{Q}_{a}(0)^{-1}Q_{a-1}(i/2)Q_{a+1}(i/2)\label{eq:QtildeQ}
\end{equation}

Let us start from the overlap (\ref{eq:start}) and perform a duality
on $Q_{3}$. Using (\ref{eq:QtildeQ}) we obtain 
\begin{equation}
S_{2}^{(3,1,4,2)}\sim\frac{Q_{1}(0)}{Q_{2}(0)\tilde{Q}_{3}(0)}\frac{\det\tilde{G}^{+}}{\det\tilde{G}^{-}}\longrightarrow\frac{Q_{1}(0)}{Q_{2}(0)Q_{3}(0)}\frac{\det G^{+}}{\det G^{-}}.
\end{equation}
Since we keep track of the grading in the notation it would be confusing
to carry the tilde over the duality (and further dualities) thus we
suppress it in the notation. Each quantity, $Q$ , ${\rm det}G^{\pm}$
is understood in the respective grading.

A further duality at $Q_{1}$ provides the sought for overlap in the
grading $(1,3,4,2)$
\begin{equation}
S_{2}^{(1,3,4,2)}\sim\frac{Q_{2}(\frac{i}{2})}{Q_{1}(0)Q_{2}(0)Q_{3}(0)}\frac{\det G^{+}}{\det G^{-}}.\label{eq:S2_1342}
\end{equation}
We can also reach by dualities the overlap in the grading $(1,2,3,4)$
\begin{equation}
S_{2}^{(1,2,3,4)}\sim\frac{1}{Q_{1}(0)Q_{1}(i/2)}\frac{Q_{2}(0)}{1}\frac{Q_{3}(0)}{Q_{3}(i/2)}\frac{\det G^{+}}{\det G^{-}}
\end{equation}
which agrees with the overlap formulas of the $su(2)$ subsectors.
All of these overlaps have factorized forms but they might not correspond
to Bethe states in the new grading. In the Appendix we investigate
carefully in which gradings the boundary state can have overlaps with
Bethe states and calculate carefully the scalar coefficients.

\subsection{Non-invertible K-matrices and overlaps of descendant states}

Let us now investigate the non-invertible K-matrices, since they appear
in the relevant applications. They again come in two types 
\begin{equation}
K^{(1)}=\left(\begin{array}{cccc}
k_{1} & k_{2}+\frac{s}{u+i/2} & 0 & 0\\
k_{2}-\frac{s}{u+i/2} & k_{4} & 0 & 0\\
0 & 0 & 0 & 0\\
0 & 0 & 0 & 0
\end{array}\right),\quad K^{(2)}=\left(\begin{array}{cccc}
0 & 0 & 0 & 0\\
0 & 0 & 0 & 0\\
0 & 0 & k_{1} & k_{2}+\frac{s}{u+i/2}\\
0 & 0 & k_{2}-\frac{s}{u+i/2} & k_{4}
\end{array}\right),
\end{equation}
which are the weak coupling versions of the $K$-matrices when $x_{s}$
is $g$-dependent. Since in the $g\to0$ limit we obtained the overlap
for the first type of K-matrix 
\begin{equation}
S_{1}^{(1,3,4,2)}\sim\frac{Q_{1}(is)^{2}Q_{3}(0)}{Q_{1}(0)Q_{2}(0)Q_{2}(\frac{i}{2})}\frac{\det G^{+}}{\det G^{-}}
\end{equation}
we can use the transformation $1\leftrightarrow3$; $2\leftrightarrow4$
to obtain the overlap of the second type of K-matrix in the opposite
grading
\begin{equation}
S_{2}^{(3,1,2,4)}\sim\frac{Q_{1}(is)^{2}Q_{3}(0)}{Q_{1}(0)Q_{2}(0)Q_{2}(\frac{i}{2})}\frac{\det G^{+}}{\det G^{-}}.
\end{equation}
which has a factorized form. We now would like to calculate this overlap
in the original grading $(1,3,4,2)$. This amounts to use the duality
transformation for $Q_{1}$: 
\begin{equation}
S_{2}^{(1,3,2,4)}\sim\frac{Q_{1}(0)\tilde{Q}_{1}(is)^{2}Q_{3}(0)}{Q_{2}(0)Q_{2}(\frac{i}{2})^{2}}\frac{\det G^{+}}{\det G^{-}}.
\end{equation}
 written in terms of the new $Q_{1}(0).$ But then the previous $Q_{1}(is)$
term, denoted by $\tilde{Q}_{1}(is)$ in this grading, is problematic
as it should be expanded by the $QQ$-relations $\tilde{Q}_{1}\sim Q_{1}^{-1}(Q_{\theta}^{+}Q_{2}^{-}-Q_{\theta}^{-}Q_{2}^{+})$
used at $s\neq0$ implying that the overlap $S_{2}^{(1,3,2,4)}$ does
not have a factorized form
\begin{equation}
S_{2}^{(1,3,2,4)}\sim\frac{Q_{1}(0)Q_{1}^{-1}((Q_{\theta}^{+}(is)Q_{2}^{-}(is)-Q_{\theta}^{-}(is)Q_{2}^{+}(is))^{2}Q_{3}(0)}{Q_{1}(is)^{2}Q_{2}(0)Q_{2}(\frac{i}{2})^{2}}\frac{\det G^{+}}{\det G^{-}}.
\end{equation}

This seems to contradict with the assumptions in \cite{Kristjansen:2020vbe},
that overlaps of Bethe states have factorized forms. The way out is
what we explained, that the Bethe state under duality transforms to
a descendant state. If both the Bethe state and its descendants have
non-zero overlaps then they differ only by some constant factors,
what we typically omit. If, however, the highest weight Bethe state
has a zero overlap, and only the descendants overlap, then this overlap
is not necessarily factorizing.

For $K^{(2)}$ the non-vanishing overlaps require the selection rule
$\#1=\#2=0$. From considerations in Appendix \ref{sec:App_desc}
it follows that there are no Bethe states with non-vanishing overlaps
for the gradings $(1,2,3,4)$, $(1,3,4,2)$, $(1,3,2,4)$, $(3,1,2,4)$
and $(3,1,4,2)$ therefore only descendant states can have non-vanishing
overlaps there. For the grading $(3,4,1,2)$ Bethe states can have
non-trivial overlaps. This overlap is non-vanishing only in the $\mathfrak{su}(2)$
sector when $N_{2}=N_{3}=0$. Since we know the $\mathfrak{su}(2)$
overlaps, we can write that 
\begin{equation}
S_{2}^{(3,4,1,2)}\sim\frac{Q_{1}(is)^{2}}{Q_{1}(0)Q_{1}(\frac{i}{2})}\frac{\det G^{+}}{\det G^{-}}.
\end{equation}
This grading is very special for the boundary state corresponding
to $K^{(2)}$ since this is the only one, when there are Bethe-states
with non-vanishing overlaps. The corresponding overlap has actually
a factorized form. For gradings where only descendant states can have
non-vanishing overlaps, the overlap does not necessarily factorize.
Indeed, for the grading $(3,1,2,4)$ it has but for $(1,2,3,4)$ or
$(1,3,4,2)$ it does not.

In Appendix \ref{sec:App_desc} we investigate explicitly the relations
between Bethe states in one grading and descendant states in other
gradings. Besides determining the requirement for non-vanishing overlaps
we also demonstrate how the precise overlaps for descendant states
can be calculated. This amounts to use heavily the $su(2|2)$ symmetry
algebra (\ref{eq:su2|2}) and the $osp(2|2)$ symmetry of the boundary
state (\ref{eq:E23E24overlap-1}).

We close the section by summarizing what we have learned so far.
\begin{itemize}
\item The overlaps corresponding to the two types of $K$-matrices (\ref{eq:ratK})
can be described by the same formulae, but they belong to opposite
gradings (\ref{eq:idOV}).
\item In order to obtain the overlaps of the other type of K-matrix in the
original grading, or other gradings, we can use the fermionic duality
formulae. These formulae, however, in special gradings can lead to
non-factorizing overlaps, signaling that the overlap of the Bethe
state vanishes. There are special gradings where all Bethe states
have vanishing overlaps and only descendants state can overlap with
the boundary state.
\item For descendant states the overlaps can be expressed in terms of the
overlaps with Bethe states in another grading. In doing so one has
to use the duality formulas. together with the connection between
Bethe states (\ref{eq:conn}), the $osp(2|2)$ symmetry of the boundary
state (\ref{eq:E23E24overlap-1}) and the $su(2|2)$ symmetry algebra
(\ref{eq:su2|2}).
\end{itemize}
We are going to face similar situation for the centrally extended
$su(2\vert2)_{c}$ algebra and the corresponding K-matrices. Before
turning to this situation we extend the previous analysis for $gl(4\vert4)$
overlaps, which appear at the weak coupling limit of the AdS/dCFT
correspondence.

\section{Overlaps and dualities for rational $gl(4|4)$ spin chains, the leading
order 't Hooft line}

The weak coupling limit of the integrable description governing the
$AdS_{5}/CFT_{4}$ duality can be described by the $gl(4\vert4)$
spin chain \cite{Beisert:2005fw}. In the following we lift the previous
result for this case. The bosonic part of the $gl(4|4)$ K-matrices
are diagonal sums of the form 
\begin{equation}
\mathbb{K}=K^{(+)}\oplus K^{(-)}.
\end{equation}
As for the $su(2\vert2)$ case, there are two solutions of the bYBE,
which can be distinguished how they transform for transposition in
the respective sub-spaces. For the first solution $\mathbb{K}^{(1)}$:
$\left(K^{(1),(\pm)}\right)^{t}=\pm K^{(1),(\pm)}$, while for the
second $\mathbb{K}^{(2)}$: $\left(K^{(2),(\pm)}\right)^{t}=\mp K^{(2),(\pm)}$.
In the $AdS_{5}/CFT_{4}$ context the $(+)$ bosonic subspace corresponds
to the isometries of $S^{5}$ while the $(-)$ bosonic subspace to
the isometries of $AdS_{5}$. The boundary related to $\mathbb{K}^{(1)}$
breaks the isometries of $S^{5}$ and $AdS_{5}$ to $SO(3)\times SO(3)$
and $SO(2,3)$, respectively, which is the bosonic symmetry of the
D3-D5 domain wall defect. Contrary, the boundary related to $\mathbb{K}^{(2)}$
breaks the isometries of $S^{5}$ and $AdS_{5}$ to $SO(5)$ and $SO(2,1)\times SO(3)$,
respectively, which is the symmetry of the 't Hooft line.

Since we have already obtained the weak coupling asymptotic overlaps
for $\mathbb{K}^{(1)}$, our goal is to derive the overlaps corresponding
to the other $K$-matrix $\mathbb{K}^{(2)}$ and demonstrate that
these overlaps indeed describe the 't Hooft line at weak coupling.
We are going to achieve this goal by exploiting the connection between
the $K$-matrices $K^{(1),(\pm)}=K^{(2),(\mp)}$ and using various
duality transformations. Let us denote the overlap $S_{k}^{(\theta_{1}\theta_{2}\theta_{3}\theta_{4}\theta_{5}\theta_{6}\theta_{7}\theta_{8})}$
for the $K$-matrix $\mathbb{K}^{(k)}$ in the grading $\theta_{j}$
($\theta_{j}=\pm$). By using the connection between the $K$-matrices
we have the relation
\begin{equation}
S_{2}^{(\theta_{1}\theta_{2}\theta_{3}\theta_{4}\theta_{5}\theta_{6}\theta_{7}\theta_{8})}=S_{1}^{(-\theta_{1},-\theta_{2},-\theta_{3},-\theta_{4},-\theta_{5},-\theta_{6},-\theta_{7},-\theta_{8})}.\label{eq:S1S2weakfull}
\end{equation}

We start with duality transformations for the first type of overlaps
$S_{1}$. For the alternating Dynkin diagram $\otimes-\Circle-\otimes-\Circle-\otimes-\Circle-\otimes$
there are two gradings $(+--++--+)$ and $(-++--++-)$, in which the
massive particles (related to $Q_{4}$) correspond to the $SU(2)$
or $SL(2)$ sectors, respectively. For the grading $(+--++--+)$ and
the $K$-matrix $\mathbb{K}^{(1)}$, the weak coupling overlap is
\begin{equation}
S_{1}^{(+--++--+)}=\frac{Q_{1}Q_{3}Q_{4}Q_{5}Q_{7}}{Q_{2}Q_{2}^{+}Q_{4}^{+}Q_{6}Q_{6}^{+}}\frac{\det G^{+}}{\det G^{-}}.
\end{equation}
where $Q_{j}=Q_{j}(0)$, $Q_{j}^{+}=Q_{j}(\frac{i}{2})$ denote the
$Q$-functions in the given grading. By dualizing $Q_{3}$ we can
reach the grading with the overlap
\begin{equation}
S_{1}^{(+-+-+--+)}=\frac{Q_{1}Q_{4}Q_{5}Q_{7}}{Q_{2}Q_{3}Q_{6}Q_{6}^{+}}\frac{\det G^{+}}{\det G^{-}}.
\end{equation}
As we explained before the meaning of $Q_{3}$ and the ratio of determinants
is different from the previous formula and they are all understood
in their respective gradings. Performing a duality on $Q_{5}$ leads
to 
\begin{equation}
S_{1}^{(+-+--+-+)}=\frac{Q_{1}Q_{4}Q_{4}^{+}Q_{7}}{Q_{2}Q_{3}Q_{5}Q_{6}}\frac{\det G^{+}}{\det G^{-}}.
\end{equation}
Finally, dualizing $Q_{1}$ and $Q_{7}$ we obtain 
\begin{equation}
S_{1}^{(-++--++-)}=\frac{Q_{2}^{+}Q_{4}Q_{4}^{+}Q_{6}^{+}}{Q_{1}Q_{2}Q_{3}Q_{5}Q_{6}Q_{7}}\frac{\det G^{+}}{\det G^{-}},
\end{equation}
which is the overlap for the same alternating diagram, we started
with, but with the massive particle in the $SL(2)$ sector. By using
the identity, which relates the two K-matrices (\ref{eq:S1S2weakfull}),
we can obtain the overlap of the other $K$-matrix for the alternating
diagram with massive particles in the $SU(2)$ sector as 
\begin{equation}
S_{2}^{(+--++--+)}=\frac{Q_{2}^{+}Q_{4}Q_{4}^{+}Q_{6}^{+}}{Q_{1}Q_{2}Q_{3}Q_{5}Q_{6}Q_{7}}\frac{\det G^{+}}{\det G^{-}}.
\end{equation}

In order to compare this overlap with results for the 't Hooft line
we perform various duality transformations. Let us first calculate
the overlap in the grading $\Circle-\otimes-\Circle-\Circle-\Circle-\otimes-\Circle$
with $(--++++--)$. Dualizing $Q_{1}$ and $Q_{7}$ leads to 
\begin{equation}
S_{2}^{(-+-++-+-)}=\frac{Q_{1}Q_{4}Q_{4}^{+}Q_{7}}{Q_{2}Q_{3}Q_{5}Q_{6}}\frac{\det G^{+}}{\det G^{-}}.
\end{equation}
By dualizing further $Q_{2}$ and $Q_{6}$ we obtain
\begin{equation}
S_{2}^{(--++++--)}=\frac{Q_{1}Q_{2}Q_{4}Q_{4}^{+}Q_{6}Q_{7}}{Q_{1}^{+}Q_{3}Q_{3}^{+}Q_{5}Q_{5}^{+}Q_{7}^{+}}\frac{\det G^{+}}{\det G^{-}}.
\end{equation}
We can project this result into the $SO(6)$ subsector by switching
off other excitations. The overlap is simply
\begin{equation}
S_{2}^{(--++++--)}=\frac{Q_{4}Q_{4}^{+}}{Q_{3}Q_{3}^{+}Q_{5}Q_{5}^{+}}\frac{\det G^{+}}{\det G^{-}}.
\end{equation}
which completely agrees with the $SO(5)$ overlap formula in the $SO(6)$
sector of the 't Hooft line \cite{Kristjansen:2023ysz}.

By dualizing further we obtain that
\begin{equation}
S_{2}^{(++----++)}=\frac{Q_{2}Q_{3}Q_{4}Q_{5}Q_{6}}{Q_{1}Q_{1}^{+}Q_{3}^{+}Q_{4}^{+}Q_{5}^{+}Q_{7}Q_{7}^{+}}\frac{\det G^{+}}{\det G^{-}},
\end{equation}
which in the $SL(2)$ subsector leads to 
\begin{equation}
S_{2}^{(++----++)}=\frac{Q_{4}}{Q_{4}^{+}}\frac{\det G^{+}}{\det G^{-}},
\end{equation}
in complete agreement with \cite{Kristjansen:2023ysz}.

Performing one last duality leads to 
\begin{equation}
S_{2}^{(----++++)}=\frac{Q_{1}Q_{2}Q_{3}Q_{4}Q_{6}Q_{6}^{+}}{Q_{1}^{+}Q_{2}^{+}Q_{3}^{+}Q_{5}Q_{5}^{+}Q_{7}Q_{7}^{+}}\frac{\det G^{+}}{\det G^{-}}
\end{equation}
which provides the overlap in the gluon subsector 
\begin{equation}
S_{2}^{(----++++)}=\frac{Q_{3}}{Q_{3}^{+}}\frac{\det G^{+}}{\det G^{-}}
\end{equation}
agreeing with \cite{Kristjansen:2023ysz}.

In conclusion, we have determined the overlaps of the second type
of K-matrix, $\mathbb{K}^{(2)}$ in the various gradings and demonstrated
that they reproduce all the overlaps which were available for the
weak coupling limits of the 't Hooft line. Motivated by this matching
we calculate the overlaps of the second type of K-matrix for the $su(2\vert2)_{c}$
spin chains in the next section.

\section{Overlaps and dualities for the $su(2|2)_{c}$ spin chain}

In the $su(2\vert2)_{c}$ spin chain there are only two non-equivalent
gradings. In the first the pseudovacuum is a bosonic tensor-product
state $|0^{b}\rangle=|1\rangle^{\otimes2L}$ from which the $y$ roots
create a fermionic label, say $3$ and then the $w$ root another
fermionic one with $4$. Due to the $34\to12$ scattering process,
states with label $2$ are automatically created. We can introduce
the bosonic and fermionic quantum numbers $n_{b}=\#1-\#2$, $n_{f}=\#3-\#4,$
such that the corresponding Bethe states $|\mathbf{y},\mathbf{w}\rangle^{b}$
have the quantum numbers $n_{b}=2L-N$ and $n_{f}=N-2M$. We have
obtained the normalized overlap in this grading for the first type
of $K$-matrix as
\begin{equation}
S_{1}^{b}(\mathbf{y},\mathbf{w})=k_{1}^{2L-N}x_{s}^{-N}\frac{\mathcal{R}_{y}(x_{s})^{2}}{\mathcal{R}_{y}(0)}\frac{1}{Q_{w}(0)Q_{w}(\frac{i}{2g})}\frac{\det G^{+}}{\det G^{-}},
\end{equation}
where $N=2M$.

Alternatively, we could use the fermionic pseudo vacuum $|0^{f}\rangle=|3\rangle^{\otimes2L}$
to create Bethe state, by flipping fermions to bosons first, and then
bosons to bosons. Actually the transformation $1\leftrightarrow3,2\leftrightarrow4$
not only changes the gradings but also exchanges the two K-matrices.
This implies that the overlap corresponding to the other $K$-matrix
$K^{(2)}$ can be described by the same overlap formula in the other
grading, when the Bethe state is created from the fermionic vacuum
\begin{equation}
S_{2}^{f}(\mathbf{y},\mathbf{w})=S_{1}^{b}(\mathbf{y},\mathbf{w}),
\end{equation}
thus
\begin{equation}
S_{2}^{f}(\mathbf{y},\mathbf{w})=k_{1}^{2L-N}x_{s}^{-N}\frac{\mathcal{R}_{y}(x_{s})^{2}}{\mathcal{R}_{y}(0)}\frac{1}{Q_{w}(0)Q_{w}(\frac{i}{2g})}\frac{\det G^{+}}{\det G^{-}},
\end{equation}
where again, the various $Q$-functions and determinants are understood
in their own grading. This could be the result for the overlaps in
the case of the 't Hooft line for any coupling, when wrapping effects
are neglected. We, however, would like to calculate the same overlap
in the bosonic grading. In doing so we need to implement the duality
transformations \cite{Beisert:2005if}.

\subsection{Duality relations}

Let us dualize the $y$-roots in order to describe the transfer matrix
eigenvalue in the other grading. We start from the bosonic grading
and the Bethe state $|\mathbf{y},\mathbf{w}\rangle^{b}$. The idea
is to investigate the quantity, which is related to the $y$-type
Bethe ansatz equation $(\mathcal{R}^{(-)}Q_{w}^{-}-{\cal R}^{(+)}Q_{w}^{+})$.
By definition it vanishes at $x=y_{j}$. Let us calculate the leading
orders of this expression
\begin{equation}
\mathcal{R}^{(-)}Q_{w}^{-}-{\cal R}^{(+)}Q_{w}^{+}=0\times x^{2L+M}+\sum_{c=-M+1}^{2L+M-1}c_{k}x^{k}+\left(\mathcal{R}^{(-)}(0)-\mathcal{R}^{(+)}(0)\right)\times x^{-M+1}
\end{equation}
Since $\mathcal{R}^{(-)}(0)=\mathcal{R}^{(+)}(0)$ the expression
$x^{M-1}(\mathcal{R}^{(-)}Q_{w}^{-}-{\cal R}^{(+)}Q_{w}^{+})$ is
a polynomial of order $2L+2M-2$ and it has zeros at $x=y_{j}$. Let
us denote the remaining zeros by $\tilde{y}_{j}$ and their generating
function by $\tilde{{\cal R}}_{y}$: 
\begin{equation}
\tilde{{\cal R}}_{y}=\prod_{j=1}^{\tilde{N}}(x(u)-\tilde{y}_{j})
\end{equation}
They are defined by 
\begin{equation}
x^{M-1}(\mathcal{R}^{(-)}Q_{w}^{-}-{\cal R}^{(+)}Q_{w}^{+})=A{\cal R}_{y}\tilde{{\cal R}}_{y}\label{eq:key}
\end{equation}
where $A$ is a constant and by comparing the degrees we can see that
$\tilde{N}=2L+2M-N-2.$ We have also a similar equation for the ${\cal B}$
quantities:
\begin{equation}
\frac{1}{x^{M-1}}(\mathcal{B}^{(-)}Q_{w}^{-}-{\cal B}^{(+)}Q_{w}^{+})=A{\cal B}_{y}\tilde{{\cal B}}_{y}.
\end{equation}
The dual Bethe roots $\tilde{y}$ satisfies the dual Bethe equations.
They can be obtained by evaluating (\ref{eq:key}) at $x=\tilde{y}_{j}$
leading to 
\begin{equation}
\left\{ \frac{\mathcal{R}^{(-)}}{\mathcal{R}^{(+)}}-\frac{Q_{w}^{+}}{Q_{w}^{-}}\right\} _{x(u)=\tilde{y}_{j}}=0
\end{equation}
By evaluating (\ref{eq:key}) at $u\to w_{l}\pm\frac{i}{2g}$ and
taking the ratio we have the equations 
\begin{equation}
\left\{ \frac{{\cal R}_{y}^{+}}{{\cal R}_{y}^{-}}=-\frac{\tilde{{\cal R}}_{y}^{-}}{\tilde{{\cal R}}_{y}^{+}}\left(\frac{x^{+}}{x^{-}}\right)^{M-1}\frac{{\cal R}^{(+)+}Q_{w}^{++}}{\mathcal{R}^{(-)-}Q_{w}^{--}}\right\} _{u=w_{l}}
\end{equation}
 and
\begin{equation}
\left\{ \frac{{\cal B}_{y}^{+}}{{\cal B}_{y}^{-}}=-\frac{\tilde{{\cal B}}_{y}^{-}}{\tilde{{\cal B}}_{y}^{+}}\left(\frac{x^{-}}{x^{+}}\right)^{M-1}\frac{{\cal B}^{(+)+}Q_{w}^{++}}{\mathcal{B}^{(-)-}Q_{w}^{--}}\right\} _{u=w_{l}}
\end{equation}
This implies that 
\[
\left\{ \frac{Q_{w}^{--}}{Q_{w}^{++}}\frac{\mathcal{R}_{y}^{+}\mathcal{B}_{y}^{+}}{\mathcal{R}_{y}^{-}\mathcal{B}_{y}^{-}}=\frac{\tilde{{\cal R}}_{y}^{-}}{\tilde{{\cal R}}_{y}^{+}}\frac{\tilde{{\cal B}}_{y}^{-}}{\tilde{{\cal B}}_{y}^{+}}\frac{Q_{w}^{++}}{Q_{w}^{--}}=-1\right\} _{u=w_{l}}
\]
meaning, that the dual roots satisfy the same Bethe equations as the
original ones but they correspond to different solutions. The corresponding
Bethe states $|\tilde{\mathbf{y}},\mathbf{w}\rangle^{f}$ have the
quantum numbers $n_{b}=\tilde{N}-2M=2L-N-2$ and $n_{f}=2L-\tilde{N}=N-2M+2$,
therefore the Bethe states $|\tilde{\mathbf{y}},\mathbf{w}\rangle^{f}$
and $|\mathbf{y},\mathbf{w}\rangle^{b}$ are not equal but they are
in the same $\mathfrak{su}(2|2)_{c}$ multiplet and they are connected
as
\begin{equation}
|\tilde{\mathbf{y}},\mathbf{w}\rangle^{f}=\mathbb{Q}_{3}^{\:1}\mathbb{Q}_{2}^{\dagger\,4}|\mathbf{y},\mathbf{w}\rangle^{b}
\end{equation}
where we used the standard notation for the $su(2|2)_{c}$ generators,
see Appendix \ref{sec:Appsu22c} for details\textcolor{red}{.}

\subsection{Dual overlap formulae}

We now derive the dual overlap formulae. At first let us check the
selection rules. From the form of the $K$-matrix $K^{(2)}$ we can
see that the non-vanishing overlap requires $n_{b}=0$, i.e. $2L=N$.
However, Bethe states could contain maximum $2L-2$ $y$-roots, therefore
they must have vanishing overlaps and only descendant overlaps can
be non-zero. Let us denote a descendant of the Bethe state $\vert\mathbf{y},\mathbf{w}\rangle$
by $\vert\mathbf{y},\mathbf{w},d\rangle$ and the corresponding overlap
by $S_{2}^{b}(\mathbf{y},\mathbf{w},d)$. All these descendants are
in the same $su(2|2)_{c}$ multiplet as $|\tilde{\mathbf{y}},\mathbf{w}\rangle^{f}$.
It implies that their overlaps are proportional to $S_{2}^{f}(\tilde{\mathbf{y}},\mathbf{w})$:
\begin{equation}
S_{2}^{b}(\mathbf{y},\mathbf{w};d)=C_{d}S_{2}^{f}(\tilde{\mathbf{y}},\mathbf{w})=C_{d}k_{1}^{2L-N}x_{s}^{-N}\frac{\tilde{\mathcal{R}}_{y}(x_{s})^{2}}{\mathcal{\tilde{R}}_{y}(0)}\frac{1}{Q_{w}(0)Q_{w}(\frac{i}{2g})}\frac{\det\tilde{G}^{+}}{\det\tilde{G}^{-}},\label{eq:dualoverlap}
\end{equation}
$d$ refers to the specific descendant state and $C_{d}$ is its corresponding
proportionality factor to be fixed.

In expressing the overlap in terms of bosonic quantities we start
with the ratio of determinants. We investigated this quantity in many
cases numerically and observed that 
\begin{equation}
\frac{\det\tilde{G}^{+}}{\det\tilde{G}^{-}}=\pm\frac{\det G^{+}}{\det G^{-}}
\end{equation}
was always satisfied. The sign difference is irrelevant when we compare
squares of overlaps, which is natural as the phases of Bethe states
are conventional. The remaining part of the dual overlap formula (\ref{eq:dualoverlap})
is still not satisfactory as it contains the dual Bethe roots, $\mathcal{\tilde{R}}_{y}$.
We would like to express the overlap with the original Bethe roots
only. In doing so we use the identity
\begin{equation}
{\cal R}_{y}(x_{s})\tilde{{\cal R}}_{y}(x_{s})=\frac{x_{s}^{M-1}}{A}(\mathcal{R}^{(-)}(x_{s})Q_{w}^{-}(s)-{\cal R}^{(+)}(x_{s})Q_{w}^{+}(s))
\end{equation}
where the prefactor $A$ can be obtained by evaluating the identity
at zero:
\begin{equation}
{\cal R}_{y}(0)\tilde{{\cal R}}_{y}(0)={\cal R}^{(+)}(0)\frac{\frac{i}{g}(2L-2M)-A}{A},\quad A=\left[2\sum_{j=1}^{L}\left(x_{j}^{+}-x_{j}^{-}\right)-M\frac{i}{g}\right]
\end{equation}
The overlap then can be written in terms of purely bosonic quantities
as 
\begin{multline}
S_{2}^{b}(\mathbf{y},\mathbf{w};d)=C_{d}k_{1}^{2L-N}x_{s}^{2M-N-2}\left(\frac{\mathcal{R}^{(-)}(x_{s})Q_{w}^{-}(s)-{\cal R}^{(+)}(x_{s})Q_{w}^{+}(s)}{{\cal R}_{y}(x_{s})}\right)^{2}\times\\
\frac{\mathcal{R}_{y}(0)}{{\cal R}^{(+)}(0)A(\frac{i}{g}(2L-2M)-A)}\frac{1}{Q_{w}(0)Q_{w}(\frac{i}{2g})}\frac{\det G^{+}}{\det G^{-}}.\label{eq:Sb2}
\end{multline}
Clearly, this is not a factorized overlap which is related to the
fact that only the descendants can overlap with the boundary state
in this grading. There is only one example when the overlap is factorized
which is the $s=0$ case when
\begin{equation}
S_{2}^{b}(\mathbf{y},\mathbf{w};d)=C_{d}k_{1}^{2L-N}x_{s}^{2M-N-2}\frac{({\cal R}^{(+)}(i)-(-1)^{M}\mathcal{R}^{(-)}(i))^{2}}{{\cal R}^{(+)}(0)A(\frac{i}{g}(2L-2M)-A)}\frac{\mathcal{R}_{y}(0)}{{\cal R}_{y}(i)^{2}}\frac{Q_{w}(\frac{i}{2g})}{Q_{w}(0)}\frac{\det G^{+}}{\det G^{-}}.
\end{equation}

We can also take the $g\to0$ limit. Assuming that the unfixed parameter
$x_{s}$ is regular in the $g\to0$ limit (i.e. $x_{s}=a+\mathcal{O}(g)$
where $a\neq0$) we obtain that
\begin{equation}
\lim_{g\to0}S_{2}^{b}(\mathbf{y},\mathbf{w};d)\sim\frac{1}{Q_{1}Q_{3}}\frac{Q_{2}^{+}}{Q_{2}}\frac{\det G^{+}}{\det G^{-}}.
\end{equation}
which agrees with our previous result (\ref{eq:S2_1342}).

If we would like to calculate also the normalization constants (which
is necessary when we want to compare the formulas with the actual
overlaps) then we need to know the relation between the descendant
state with non-vanishing overlap and the dual state. The simplest
case happens when $N=2L-2$, since then the descendent state can be
\begin{equation}
|\mathbf{y},\mathbf{w};-2,2\rangle^{b}:=\mathbb{Q}_{3}^{\:1}\mathbb{Q}_{2}^{\dagger\,4}|\mathbf{y},\mathbf{w}\rangle^{b}
\end{equation}
which is just $|\tilde{\mathbf{y}},\mathbf{w}\rangle^{f}$ implying
that $C_{-2,2}=1$. Here the descendent is labelled by its $n_{b}$
and $n_{f}$ quantum numbers. Clearly both $\mathbb{Q}_{3}^{1}$ and
$\mathbb{Q}_{2}^{\dagger4}$ decreases $n_{b}$ and increases $n_{f}$
by $1$.

Let us see an other example for a descendant of $N=2L-2$ with a non-vanishing
overlap
\begin{equation}
|\mathbf{y},\mathbf{w};-2,0\rangle^{b}:=\mathbb{Q}_{3}^{\:1}\mathbb{Q}_{4}^{\:1}|\mathbf{y},\mathbf{w}\rangle^{b}
\end{equation}
In this case we also need to use the symmetry properties of the boundary
state and have to calculate the norm of the descendant state. The
calculations are relegated to Appendix \ref{sec:Appsu22c}. The final
result is that $C_{-2,0}=\left(\frac{x_{s}}{k_{1}}\right)^{2}\frac{(igA+2L-2M)}{(igA+N-2M+1)}$.
In Appendix \ref{sec:Appsu22c} we provide more examples how to calculate
the normalizations properly.

We have tested these formulae numerically very extensively for various
spin chain sizes and randomly chosen rapidity parameters, so we are
quite convinced about their correctness.

\section{All loop overlap for $\mathbf{\mathbb{K}}^{(2)}$, the asymptotic
't Hooft loop}

In this section we present formulae for the asymptotic overlaps with
$su(2\vert2)_{c}\otimes su(2\vert2)_{c}$ symmetry. We analyze the
boundary K-matrix of the form 
\begin{equation}
\mathbb{K}^{(2)}(p)=K_{0}(p)K^{(2)}(p)\otimes K^{(2)}(p)
\end{equation}
where $K_{0}(p)$ can be fixed from unitarity, boundary crossing unitarity
and additional physical requirements. Different $K_{0}(p)$ factors
correspond to different physical situations. We will not distinguish
them and keep $K_{0}(p)$ in its abstract form.

The large volume spectrum of the AdS5/CFT4 correspondence is described
by the asymptotical Bethe ansatz based on the factorizing S-matrix
(\ref{eq:Smatrix}). We put $2L$ particles in a finite volume $J$
(the R-charge) and demand the periodicity of the wave function. This
problem reduces to the diagonalization of a tensor product of two
$su(2\vert2)_{c}$ transfer matrices $t(u)\otimes t(u)$. They can
be done separately for the two copies. We distinguish the left and
right factors by an upper index $\alpha=1,2$: 
\begin{equation}
t(u)\vert\mathbf{y}^{(\alpha)},\mathbf{w}^{(\alpha)}\rangle=\Lambda(u,\mathbf{y}^{(\alpha)},\mathbf{w}^{(\alpha)})\vert\mathbf{y}^{(\alpha)},\mathbf{w}^{(\alpha)}\rangle
\end{equation}
They satisfies the same Bethe equations but are typically different
solutions. The quantization of the momenta requires 
\begin{equation}
e^{\phi_{p_{j}}}:=e^{ip_{j}J}\prod_{k:k\neq j}S_{0}(p_{j},p_{k})\Lambda(u_{j},\mathbf{y}^{(1)},\mathbf{w}^{(1)})\Lambda(u_{j},\mathbf{y}^{(2)},\mathbf{w}^{(2)})=1
\end{equation}

In order to have a non-trivial overlap, momenta should come in pairs
$\mathbf{p}=\{\mathbf{p}^{+},\mathbf{p}^{-}\}$, such that $\mathbf{p}^{+}=-\mathbf{p}^{-}$
and $J=2L$. The overlap of the boundary state built from the tensor
product $\mathbb{K}$ matrix factorizes into two $su(2\vert2)_{c}$
copies. We should decide again in which grading we are interested
in the formulas. The simplest is choosing fermionic grading in both
wings, which leads to 
\begin{align}
\frac{\left|\left\langle \Psi_{\mathbb{K}^{(2)}}\right|\left.\mathbf{p},\mathbf{y}^{(\alpha)},\mathbf{w}^{(\alpha)}\right\rangle \right|^{2}}{\left\langle \mathbf{p},\mathbf{y}^{(\alpha)},\mathbf{w}^{(\alpha)}\right.\left|\mathbf{p},\mathbf{y}^{(\alpha)},\mathbf{w}^{(\alpha)}\right\rangle }=\prod_{i=1}^{L}\left|K_{0}(p_{i})\right|^{2} & \mathcal{\bar{S}}_{2}^{f}(\mathbf{y}^{(1)},\mathbf{w}^{(1)})\mathcal{\bar{S}}_{2}^{f}(\mathbf{y}^{(2)},\mathbf{w}^{(2)})\frac{{\rm det}G^{+}}{{\rm det}G^{-}}
\end{align}
where
\begin{equation}
\mathcal{\bar{S}}_{2}^{f}(\mathbf{y},\mathbf{w})=k_{1}^{2L-N}x_{s}^{-N}\frac{\mathcal{R}_{y}(x_{s})^{2}}{\mathcal{R}_{y}(0)}\frac{1}{Q_{w}(0)Q_{w}(\frac{i}{2g})},
\end{equation}
and the determinants involve also differentiation wrt. the momenta:
\begin{equation}
G_{ij}^{\pm}=\left(\partial_{\hat{U}_{i}^{+}}\phi_{U_{j}^{+}}\pm\partial_{\hat{U}_{i}^{+}}\phi_{U_{j}^{-}}\right)
\end{equation}
Here $\mathbf{U}^{+}=\left\{ \mathbf{p}^{+},\mathbf{v}^{(\alpha)+},\mathbf{w}^{(\alpha)+}\right\} $
collects all the variables and $\mathbf{\hat{U}}=\left\{ \mathbf{u}^{+},\hat{\mathbf{u}}^{(\alpha)+},\hat{\mathbf{w}}^{(\alpha)+}\right\} $
is the collection of the properly normalized rapidities \cite{Gombor:2020kgu}.

If, however, we are interested in the overlaps in the bosonic gradings
then we have to use the bosonic overlap formulas for descendant states
(\ref{eq:Sb2})
\begin{align}
\frac{\left|\left\langle \Psi_{\mathbb{K}^{(2)}}\right|\left.\mathbf{p},\mathbf{y}^{(\alpha)},\mathbf{w}^{(\alpha)};d^{(\alpha)}\right\rangle \right|^{2}}{\left\langle \mathbf{p},\mathbf{y}^{(\alpha)},\mathbf{w}^{(\alpha)};d^{(\alpha)}\right.\left|\mathbf{p},\mathbf{y}^{(\alpha)},\mathbf{w}^{(\alpha)};d^{(\alpha)}\right\rangle }={\cal A}\prod_{i=1}^{L}\left|K_{0}(p_{i})\right|^{2} & \mathcal{\bar{S}}_{2}^{b}(\mathbf{y}^{(1)},\mathbf{w}^{(1)})\mathcal{\bar{S}}_{2}^{b}(\mathbf{y}^{(2)},\mathbf{w}^{(2)})\frac{{\rm det}G^{+}}{{\rm det}G^{-}}
\end{align}
which do not have a factorized form. Here $\mathcal{A}$ is the proper
normalization corresponding to the given descendant state and
\begin{equation}
\mathcal{\bar{S}}_{2}^{b}(\mathbf{y},\mathbf{w})=\left(\frac{\mathcal{R}^{(-)}(x_{s})Q_{w}^{-}(s)-{\cal R}^{(+)}(x_{s})Q_{w}^{+}(s)}{{\cal R}_{y}(x_{s})}\right)^{2}\frac{\mathcal{R}_{y}(0)}{{\cal R}^{(+)}(0)}\frac{1}{Q_{w}(0)Q_{w}(\frac{i}{2g})}.
\end{equation}

These results are very generic and are valid for any integrable boundaries
with the specifically embedded $osp(2\vert2)\oplus osp(2\vert2)$
symmetries. In order to specify to a physical situation one has to
fix the scalar factor $K_{0}$ and determine how the parameter $x_{s}$
(which could be even different for the two factors) depend on the
parameters of the model. In the case of the 't Hooft loop we can compare
the weak coupling limit of the formulas with Section 4 and conclude
that $x_{s}$ should start at weak coupling as $x_{s}=a+O(g)$, i.e.
it does not vanish in the limit. It is a challenging task to provide
an all loop expression for $x_{s}$.

\section{Conclusions}

In this work we developed fermionic dualities for overlaps in $su(2\vert2)_{c}$
spin chains with $K$-matrices having $osp(2\vert2)_{c}$ symmetries.
These symmetries come in two versions depending on how the unbroken
symmetry is embedded into the full symmetry (\ref{eq:K12}). We have
previously calculated the overlaps for one type of $K$-matrix $K^{(1)}$
in the bosonic grading, which correspond to the D3-D5 AdS/dCFT setup.
In the present work we investigated the other embedding with $K^{(2)}$,
which has the symmetry of the 't Hooft loop. The two $K$-matrices
are related by a boson-fermion duality. As a consequence, the previous
overlap formula for $K^{(1)}$ in the bosonic grading describes the
overlap of $K^{(2)}$ in the fermionic grading. We then developed
fermionic dualities to express the overlap of the boundary state coming
from $K^{(2)}$ in bosonic gradings.

Gradings encode the information how nesting happens in the Bethe ansatz,
i.e. in which order from a given pseudo vacuum the excitations are
created. The nature of the excitations together with their scatterings
depend on the grading and the nested Bethe ansatz constructs a highest
weight Bethe state, which is the eigenvector of the transfer matrix.
Descendent states have the same eigenvalue and are created by symmetry
transformations. The same eigenvalue can be described by different
gradings. A Bethe state in one grading is a descendent state in another
grading. Fermionic dualities connect the Bethe ansatz equations in
different gradings. In order to transform an overlap formula we need
to use the $QQ$-relations together with the transformation of the
ratio of Gaudin type determinants. We elaborated these transformations
in the $su(2\vert2)$ and $su(2\vert2)_{c}$ spin chains together
with the corresponding selections rules. We observed that in certain
gradings the overlap formula does not factorize over Bethe roots.
Since after the duality the Bethe state is described by a descendent
state in the dual grading we actually calculate on overlap with a
descendent state. If the Bethe state in the dual grading has a non-trivial
overlap with the boundary state then the overlap is factorising and
it differs only by a scalar factor from the descendent overlap. If,
however, the Bethe state does not overlap with the boundary state,
then the descendant overlap is not factorising.

We investigated carefully the requirements that Bethe states overlap
with the boundary state. On the way we determined the relations between
Bethe states and descendants in the various gradings, which helped
to calculate the proper prefactors in the descendant overlaps. Here
the symmetry properties of the boundary state was crucial. Eventually,
we could describe all overlaps in all gradings in the $su(2\vert2)$
and $su(2\vert2)_{c}$ spin chains. By putting together two copies
of overlap formulas we made a proposal for overlaps in the AdS5/CFT4
settings with symmetries of the 't Hoof line. We confirmed our proposal
against all available 1 loop results in the various subsectors \cite{Kristjansen:2023ysz}.
In order to have a precise all loop description one needs to know
the scalar factor of the reflection factor $K_{0}$, together with
the explicit form of the boundary parameter $x_{s}$. The absence/presence
of boundary bound-states could help in this analysis.

As future works, it would be very nice to find $K_{0}$ and $x_{s}$,
which correspond to the 't Hooft line setting.

As our results relied only on the symmetry of the boundary $K$-matrix
it is valid for other problems with the same symmetries. In particular,
\cite{Ivanovskiy:2024vel} investigates the correlation functions
on the Coulomb branch of planar ${\cal N}=4$ SYM, where the R-symmetry
is broken to $SO(5)$ just as in the case of the 't Hoof loop. If
integrability extends to all sectors and all loops, one-point functions
could be described by our formulas, with appropriately chosen $K_{0}$
and $x_{s}$. Our overlaps are expected to be applied in the ABJM
theory, where it could describe asymptotic 3-point function \cite{Yang:2021hrl},
domain wall \cite{Kristjansen:2021abc} and Wilson loop 1-point functions
\cite{Jiang:2023cdm}.

\subsection*{Acknowledgments}

We thank Charlotte Kristjansen for the discussions and the NKFIH research
Grants K134946 and PD142929 for support. ZB thanks APCTP for the invitation
to the workshop ``Integrability, Duality and Related Topics'', Pohang,
2023, where this work started. TG was also supported by the János
Bolyai Research Scholarship of the Hungarian Academy of Science.

\appendix

\section{Bethe states and descendent states \label{sec:App_desc}}

\begin{figure}
\begin{centering}
\includegraphics[width=0.25\textwidth]{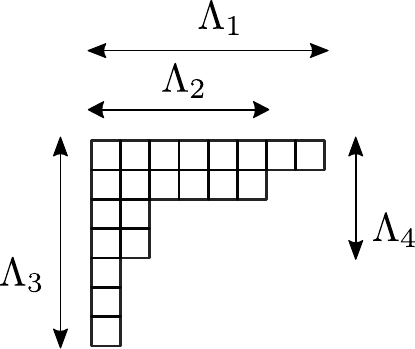}
\par\end{centering}
\caption{Young-Tableaux parametrizing the representation of $su(2\vert2)$.}

\label{YoungT}
\end{figure}

In this Appendix we investigate the relations of Bethe states in different
gradings. We start by parametrizing the irreducible representations
of $su(2|2)$ by the Young-tableaux on Figure \ref{YoungT}. The Bethe
states are highest weight states which correspond these diagrams.
They are obtained by acting with raising operators on the pseudo vacuum.
In the $\left(1,2,3,4\right)$ grading, for example, we start with
all $1$s, then use $u^{(1)}$ roots to create $2$s, then $u^{(2)}$
roots to flip $2$s to $3$s from which we flips $4$s by $u^{(3)}$.
In the corresponding state the multiplicities of $|1\rangle$, $|2\rangle$,
$|3\rangle$, $|4\rangle$ are $2L-N_{1}$, $N_{1}-N_{2},N_{2}-N_{3},N_{3}$,
respectively. They can be put in the Young-tableaux as $1$s in the
first row, $2$s in the second row, $3$s in the remaining first column
and $4$s in the last column. These are kept track by the $\Lambda$s,
as shown on Figure \ref{YoungT}. The very existence of the diagram
implies among others that $\Lambda_{1}\geq\Lambda_{2}$ and $\Lambda_{3}\geq\Lambda_{4}$.
In table \ref{Table:YT} we relate the multiplicities of $|1\rangle$,
$|2\rangle$, $|3\rangle$, $|4\rangle$ in the various gradings to
the labels of the Young-tableaux.

\begin{table}
\begin{centering}
\begin{tabular}{|c|c|c|c|c|}
\hline 
 & \#1 & \#2 & \#3 & \#4\tabularnewline
\hline 
\hline 
(1,2,3,4) & $\Lambda_{1}=2L-N_{1}$ & $\Lambda_{2}=N_{1}-N_{2}$ & $\Lambda_{3}-2=N_{2}-N_{3}$ & $\Lambda_{4}-2=N_{3}$\tabularnewline
\hline 
(1,3,4,2) & $\Lambda_{1}=2L-N_{1}$ & $\Lambda_{2}-2=N_{3}$ & $\Lambda_{3}-1=N_{1}-N_{2}$ & $\Lambda_{4}-1=N_{2}-N_{3}$\tabularnewline
\hline 
(1,3,2,4) & $\Lambda_{1}=2L-N_{1}$ & $\Lambda_{2}-1=N_{2}-N_{3}$ & $\Lambda_{3}-1=N_{1}-N_{2}$ & $\Lambda_{4}-2=N_{3}$\tabularnewline
\hline 
(3,4,1,2) & $\Lambda_{1}-2=N_{2}-N_{3}$ & $\Lambda_{2}-2=N_{3}$ & $\Lambda_{3}=2L-N_{1}$ & $\Lambda_{4}=N_{1}-N_{2}$\tabularnewline
\hline 
(3,1,2,4) & $\Lambda_{1}-1=N_{1}-N_{2}$ & $\Lambda_{2}-1=N_{2}-N_{3}$ & $\Lambda_{3}=2L-N_{1}$ & $\Lambda_{4}-2=N_{3}$\tabularnewline
\hline 
(3,1,4,2) & $\Lambda_{1}-1=N_{1}-N_{2}$ & $\Lambda_{2}-2=N_{3}$ & $\Lambda_{3}=2L-N_{1}$ & $\Lambda_{4}-1=N_{2}-N_{3}$\tabularnewline
\hline 
\end{tabular}
\par\end{centering}
\caption{Relation between the labels of the Young-tableaux and excitation numbers
in the various gradings.}

\label{Table:YT}
\end{table}

Each Young-tableaux corresponds to a given representation of $su(2\vert2)$.
For a given grading the nesting procedure provides a Bethe state,
which is a highest weight state. Other nesting uses different creation
operators and describes the same representations by different highest
weight Bethe states. A highest weight Bethe state in one grading is
a descendent state in another grading. Let us denote the Bethe state
in the grading $(A_{1},A_{2},A_{3},A_{4})$ as $|\mathbf{u}_{(A_{1},A_{2},A_{3},A_{4})}\rangle$.
Bethe states in different gradings can be connected to each other
as 
\begin{equation}
|\mathbf{u}_{(\dots,A_{k},A_{k+1},\dots)}\rangle=\frac{1}{\sqrt{\#A_{k}+\#A_{k+1}}}E_{A_{k},A_{k+1}}|\mathbf{u}_{(\dots,A_{k+1},A_{k},\dots)}\rangle,\label{eq:conn}
\end{equation}
where $E_{i,j}$ are the $su(2|2)$ generators which satisfy the relations
\begin{equation}
[E_{i,j},E_{k,l}]:=E_{i,j}E_{k,l}-(-1)^{([i]+[j])([k]+[l])}E_{k,l}E_{i,j}=\delta_{j,k}E_{i,l}-(-1)^{([i]+[j])([k]+[l])}\delta_{i,l}E_{k,j}.\label{eq:su2|2}
\end{equation}
The structure of $K^{(2)}$ implies that there is the selection rule
$\#1=\#2$. Looking at the table we can see that the Bethe states
have non-vanishing overlaps in the gradings $(1,2,3,4)$, $(3,1,2,4)$
and $(3,4,1,2)$ for the representations $\Lambda_{1}=\Lambda_{2}$.
In the gradings $(1,3,4,2)$, $(1,3,2,4)$ and $(3,1,4,2)$ however
$\Lambda_{1}\geq\Lambda_{2}$ cannot be satisfied, thus there are
no Bethe states with non-vanishing overlaps i.e. $\langle\Psi|\mathbf{u}_{(1,3,4,2)}\rangle=0$
for all Bethe states.

In other words, using the duality we can obtain overlap expressions
for every grading, however not every overlap formula corresponds to
Bethe states, some of them corresponds only to descendent states.
E.g., for the grading $(1,3,4,2)$, taking an irrep with $\Lambda_{1}=\Lambda_{2}$,
the overlap with the Bethe state is zero, but the following descendants
have non-vanishing overlaps
\begin{align}
\langle\Psi|E_{2,3}E_{2,4}|\mathbf{u}_{(1,3,4,2)}\rangle & \neq0\quad;\quad\langle\Psi|E_{3,1}E_{2,4}|\mathbf{u}_{(1,3,4,2)}\rangle\neq0.
\end{align}
These overlaps are given by the expression (\ref{eq:S2_1342}) up
to a combinatorial prefactor
\begin{equation}
\frac{|\langle\Psi|E_{2,3}E_{2,4}|\mathbf{u}_{(1,3,4,2)}\rangle|^{2}}{|E_{2,3}E_{2,4}|\mathbf{u}_{(1,3,4,2)}\rangle|^{2}}=\left(\frac{s}{k_{1}}\right)^{2}\frac{L-N_{2}}{N_{1}-N_{2}+N_{3}+1}\frac{|\langle\Psi|\mathbf{u}_{(3,1,2,4)}\rangle|^{2}}{\langle\mathbf{u}_{(3,1,2,4)}|\mathbf{u}_{(3,1,2,4)}\rangle}.
\end{equation}
which we determine in the next subsection. 

\subsection{Overlap calculations in the $su(2\vert2)$ spin chain}

Assuming that we fixed the prefactor of $S_{2}^{(3,1,2,4)}$ we can
calculate the complete overlaps in the grading $(1,3,4,2)$ using
the connection between the Bethe states of different gradings (\ref{eq:conn}),
the exchange properties of the $su(2\vert2)$ algebra and the transformation
properties of the boundary state. We start with $|\mathbf{u}_{(1,3,4,2)}\rangle=E_{1,3}E_{4,2}|\mathbf{u}_{(3,1,2,4)}\rangle$
and write
\begin{align}
\langle\Psi|E_{2,3}E_{2,4}|\mathbf{u}_{(1,3,4,2)}\rangle & =\langle\Psi|E_{2,3}E_{2,4}E_{1,3}E_{4,2}|\mathbf{u}_{(3,1,2,4)}\rangle=\langle\Psi|E_{1,3}E_{2,3}E_{2,4}E_{4,2}|\mathbf{u}_{(3,1,2,4)}\rangle\label{eq:E23E24overlap-1}\\
 & =\langle\Psi|E_{1,3}E_{2,3}(E_{2,2}+E_{4,4})|\mathbf{u}_{(3,1,2,4)}\rangle=N_{2}\langle\Psi|E_{1,3}E_{2,3}|\mathbf{u}_{(3,1,2,4)}\rangle\nonumber 
\end{align}
where we only used the $su(2|2)$ algebra and the fact that the Bethe
state is a highest weight state, i.e. $E_{2,4}|\mathbf{u}_{(3,1,2,4)}\rangle=0.$
We now can exploit the $osp(2|2)$ symmetry of the $K$-matrix and
the corresponding boundary state
\begin{equation}
\sum_{k}\langle\Psi|E_{i,k}K_{k,j}^{(2)}=(-1)^{[j]([j]+[k])}\sum_{k}K_{i,k}^{(2)}\langle\Psi|E_{j,k}\label{eq:symK-1}
\end{equation}
therefore $\langle\Psi|E_{1,3}K_{3,3}+\langle\Psi|E_{1,4}K_{3,4}=-K_{1,2}\langle\Psi|E_{3,2}$,
which reads explicitly as 
\begin{equation}
\langle\Psi|E_{1,3}=\frac{s}{k_{1}}\langle\Psi|E_{3,2}-\frac{k_{2}}{k_{1}}\langle\Psi|E_{1,4}.
\end{equation}
Substituting this relation back into eq. (\ref{eq:E23E24overlap-1})
we obtain that
\begin{align}
\langle\Psi|E_{2,3}E_{2,4}|\mathbf{u}_{(1,3,4,2)}\rangle & =\frac{s}{k_{1}}N_{2}\langle\Psi|E_{3,2}E_{2,3}|\mathbf{u}_{(3,1,2,4)}\rangle=\frac{s}{k_{1}}N_{2}(L-N_{2})\langle\Psi|\mathbf{u}_{(3,1,2,4)}\rangle.
\end{align}
which expresses the overlap of the boundary state with a descendent
state in one grading to the overlap with a highest weight state in
another grading. In order to get the normalized overlap we need to
calculate the norm of the descendent state. This can be done by using
the relations of the algebra and the relations between the different
Bethe states: 
\begin{align}
 & |E_{2,3}E_{2,4}|\mathbf{u}_{(1,3,4,2)}\rangle|^{2}=\\
 & =\langle\mathbf{u}_{(1,3,4,2)}|E_{4,2}E_{3,2}E_{2,3}E_{2,4}|\mathbf{u}_{(1,3,4,2)}\rangle\nonumber \\
 & =\langle\mathbf{u}_{(1,3,4,2)}|E_{4,2}[E_{3,2},E_{2,3}]E_{2,4}|\mathbf{u}_{(1,3,4,2)}\rangle-\langle\mathbf{u}_{(1,3,4,2)}|E_{4,2}E_{2,3}E_{3,2}E_{2,4}|\mathbf{u}_{(1,3,4,2)}\rangle\nonumber \\
 & =(N_{1}-N_{2}+N_{3}+1)\langle\mathbf{u}_{(1,3,4,2)}|E_{4,2}E_{2,4}|\mathbf{u}_{(1,3,4,2)}\rangle-\langle\mathbf{u}_{(1,3,4,2)}|E_{4,2}E_{2,3}E_{3,4}|\mathbf{u}_{(1,3,4,2)}\rangle\nonumber \\
 & =N_{2}(N_{1}-N_{2}+N_{3}+1)\langle\mathbf{u}_{(1,3,4,2)}|\mathbf{u}_{(1,3,4,2)}\rangle.\nonumber 
\end{align}
together with
\begin{align}
\langle\mathbf{u}_{(1,3,4,2)}|\mathbf{u}_{(1,3,4,2)}\rangle & =\langle\mathbf{u}_{(3,1,2,4)}|E_{2,4}E_{3,1}E_{1,3}E_{4,2}|\mathbf{u}_{(3,1,2,4)}\rangle=N_{2}(L-N_{2})\langle\mathbf{u}_{(3,1,2,4)}|\mathbf{u}_{(3,1,2,4)}\rangle
\end{align}
Collecting all the formulas we can express the overlap in the (1,3,4,2)
grading in terms of the overlap in the (3,1,2,4) grading as
\begin{equation}
\frac{|\langle\Psi|E_{2,3}E_{2,4}|\mathbf{u}_{(1,3,4,2)}\rangle|^{2}}{|E_{2,3}E_{2,4}|\mathbf{u}_{(1,3,4,2)}\rangle|^{2}}=\left(\frac{s}{k_{1}}\right)^{2}\frac{L-N_{2}}{N_{1}-N_{2}+N_{3}+1}\frac{|\langle\Psi|\mathbf{u}_{(3,1,2,4)}\rangle|^{2}}{\langle\mathbf{u}_{(3,1,2,4)}|\mathbf{u}_{(3,1,2,4)}\rangle}.
\end{equation}

\section{Overlap calculations in the $su(2\vert2)_{c}$ spin chain \label{sec:Appsu22c}}

In this Appendix we calculate the overlaps of descendant states in
the $su(2\vert2)_{c}$ spin chain. We start by calculating the properly
normalized overlap for a descendant with $N=2L-2$ with a non-vanishing
overlap
\begin{equation}
|\mathbf{y},\mathbf{w};-2,0\rangle^{b}:=\mathbb{Q}_{3}^{\:1}\mathbb{Q}_{4}^{\:1}|\mathbf{y},\mathbf{w}\rangle^{b}
\end{equation}
We use the exchange relations of the $su(2\vert2)_{c}$ algebra

\begin{align}
\left[\mathbb{R}_{a}^{\:\:b},\mathbb{J}_{c}\right] & =\delta_{c}^{b}\mathbb{J}_{a}-\frac{1}{2}\delta_{a}^{b}\mathbb{J}_{c}, & \left[\mathbb{L}_{\alpha}^{\:\:\beta},\mathbb{J}_{\gamma}\right] & =\delta_{\gamma}^{\beta}\mathbb{J}_{\alpha}-\frac{1}{2}\delta_{\alpha}^{\beta}\mathbb{J}_{\gamma},\nonumber \\
\left[\mathbb{R}_{a}^{\:\:b},\mathbb{J}^{c}\right] & =-\delta_{a}^{c}\mathbb{J}^{b}+\frac{1}{2}\delta_{a}^{b}\mathbb{J}^{c}, & \left[\mathbb{L}_{\alpha}^{\:\:\beta},\mathbb{J}^{\gamma}\right] & =-\delta_{\alpha}^{\gamma}\mathbb{J}^{\beta}+\frac{1}{2}\delta_{\alpha}^{\beta}\mathbb{J}^{\gamma},\nonumber \\
\left\{ \mathbb{Q}_{\alpha}^{\:\:a},\mathbb{Q}_{\beta}^{\:\:b}\right\}  & =\epsilon_{\alpha\beta}\epsilon^{ab}\mathbb{C}, & \left\{ \mathbb{Q}_{a}^{\dagger\:\alpha},\mathbb{Q}_{b}^{\dagger\:\beta}\right\}  & =\epsilon_{ab}\epsilon^{\alpha\beta}\mathbb{C}^{\dagger},\nonumber \\
\left\{ \mathbb{Q}_{\alpha}^{\:\:b},\mathbb{Q}_{a}^{\dagger\:\beta}\right\}  & =\delta_{\alpha}^{\beta}\mathbb{R}_{a}^{\:\:b}+\delta_{a}^{b}\mathbb{L}_{\alpha}^{\:\:\beta}+\frac{1}{2}\delta_{\alpha}^{\beta}\delta_{a}^{b}\mathbb{H}
\end{align}
where $\mathbb{C}=\mathbb{C}^{\dagger}=ig\left(e^{-i\mathbb{P}/2}-e^{i\mathbb{P}/2}\right)$,
together with the symmetry properties of the boundary state
\begin{equation}
\langle\Psi^{(2)}|\mathbb{Q}_{\alpha}^{\:a}=ix_{s}\epsilon^{ab}\eta_{\alpha\beta}\langle\Psi^{(2)}|\mathbb{Q}_{b}^{\dagger\,\beta}\quad,\quad\eta^{\alpha\beta}\langle\Psi^{(2)}|\mathbb{Q}_{\beta}^{\:a}=ix_{s}\epsilon^{ab}\langle\Psi^{(2)}|\mathbb{Q}_{b}^{\dagger\,\alpha},
\end{equation}
where
\begin{equation}
\eta_{\alpha\beta}=\left(\begin{array}{cc}
k_{1} & k_{2}\\
k_{2} & k_{4}
\end{array}\right),\quad\eta^{\alpha\beta}=\left(\begin{array}{cc}
k_{4} & -k_{2}\\
-k_{2} & k_{1}
\end{array}\right).
\end{equation}
Taking into account the $osp(2\vert2)$ invariance property of the
boundary state
\begin{equation}
k_{1}\langle\Psi^{(2)}|\mathbb{Q}_{4}^{\:1}=ix_{s}\langle\Psi^{(2)}|\mathbb{Q}_{2}^{\dagger\,4}+k_{2}\langle\Psi^{(2)}|\mathbb{Q}_{3}^{\:1}
\end{equation}
the overlap simplifies as
\begin{equation}
\langle\Psi^{(2)}|\mathbf{y},\mathbf{w};-2,0\rangle^{b}:=-\frac{ix_{s}}{k_{1}}\langle\Psi^{(2)}|\mathbb{Q}_{2}^{\dagger\,4}\mathbb{Q}_{3}^{\:1}|\mathbf{y},\mathbf{w}\rangle^{b}=\frac{ix_{s}}{k_{1}}\langle\Psi^{(2)}|\tilde{\mathbf{y}},\mathbf{w}\rangle^{f}
\end{equation}
In order to calculate the normalized overlap we need to determine
the norm, too. It can be calculated as 
\begin{align}
\langle\mathbf{y},\mathbf{w};-2,0|\mathbf{y},\mathbf{w};-2,0\rangle^{b} & =\langle\mathbf{y},\mathbf{w}|\mathbb{Q}_{1}^{\dagger\,4}\mathbb{Q}_{1}^{\dagger\,3}\mathbb{Q}_{3}^{\:1}\mathbb{Q}_{4}^{\:1}|\mathbf{y},\mathbf{w}\rangle^{b}=\\
 & =\langle\mathbf{y},\mathbf{w}|\mathbb{Q}_{1}^{\dagger\,4}(\mathbb{R}_{3}^{\:3}+\mathbb{L}_{1}^{\:1}+\frac{1}{2}\mathbb{H})\mathbb{Q}_{4}^{\:1}|\mathbf{y},\mathbf{w}\rangle^{b}-\langle\mathbf{y},\mathbf{w}|\{\mathbb{Q}_{1}^{\dagger\,4},\mathbb{Q}_{3}^{\:1}\}\mathbb{R}_{4}^{\:3}|\mathbf{y},\mathbf{w}\rangle^{b}\nonumber \\
 & =-(igA+1)\langle\mathbf{y},\mathbf{w}|\mathbb{Q}_{1}^{\dagger\,4}\mathbb{Q}_{4}^{\:1}|\mathbf{y},\mathbf{w}\rangle^{b}-\langle\mathbf{y},\mathbf{w}|\mathbb{R}_{3}^{\:4}\mathbb{R}_{4}^{\:3}|\mathbf{y},\mathbf{w}\rangle^{b}\nonumber \\
 & =-(igA+1)\langle\mathbf{y},\mathbf{w}|(\mathbb{R}_{4}^{\:4}+\mathbb{L}_{1}^{\:1}+\frac{1}{2}\mathbb{H})|\mathbf{y},\mathbf{w}\rangle^{b}-2\langle\mathbf{y},\mathbf{w}|\mathbb{R}_{3}^{\:3}|\mathbf{y},\mathbf{w}\rangle^{b}\nonumber \\
 & =igA(igA+N-2M+1)\langle\mathbf{y},\mathbf{w}|\mathbf{y},\mathbf{w}\rangle^{b}\nonumber 
\end{align}
where we used that
\[
\frac{1}{2}\mathbb{H}|\mathbf{y},\mathbf{w}\rangle^{b}=\left[2ig\sum_{k=1}^{L}(x_{k}^{-}-x_{k}^{+})-L\right]|\mathbf{y},\mathbf{w}\rangle^{b}=\left[-igA-L+M\right]|\mathbf{y},\mathbf{w}\rangle^{b}
\]
We also have
\begin{align}
\langle\tilde{\mathbf{y}},\mathbf{w}|\tilde{\mathbf{y}},\mathbf{w}\rangle^{f} & =\langle\mathbf{y},\mathbf{w}|\mathbb{Q}_{4}^{\:2}\mathbb{Q}_{1}^{\dagger\,3}\mathbb{Q}_{3}^{\:1}\mathbb{Q}_{2}^{\dagger\,4}|\mathbf{y},\mathbf{w}\rangle^{b}\\
 & =\langle\mathbf{y},\mathbf{w}|\mathbb{Q}_{4}^{\:2}(\mathbb{R}_{3}^{\:3}+\mathbb{L}_{1}^{\:1}+\frac{1}{2}\mathbb{H})\mathbb{Q}_{2}^{\dagger\,4}|\mathbf{y},\mathbf{w}\rangle^{b}-\langle\mathbf{y},\mathbf{w}|\mathbb{C}\mathbb{C}^{\dagger}|\mathbf{y},\mathbf{w}\rangle^{b}\nonumber \\
 & =igA(igA+2L-2M)\langle\mathbf{y},\mathbf{w}|\mathbf{y},\mathbf{w}\rangle^{b}\nonumber 
\end{align}
where we used that
\begin{align}
\mathbb{C}|\mathbf{y},\mathbf{w}\rangle^{b}=\mathbb{C}^{\dagger}|\mathbf{y},\mathbf{w}\rangle^{b} & =ig\sum_{k=1}^{2L}\left(\sqrt{\frac{x_{k}^{+}}{x_{k}^{-}}}-\sqrt{\frac{x_{k}^{-}}{x_{k}^{+}}}\right)|\mathbf{y},\mathbf{w}\rangle^{b}\\
 & =ig\sum_{k=1}^{L}\left(\sqrt{\frac{x_{k}^{+}}{x_{k}^{-}}}-\sqrt{\frac{x_{k}^{-}}{x_{k}^{+}}}+\sqrt{\frac{x_{k}^{-}}{x_{k}^{+}}}-\sqrt{\frac{x_{k}^{+}}{x_{k}^{-}}}\right)|\mathbf{y},\mathbf{w}\rangle^{b}=0
\end{align}
In summary, the normalization constant turns out to be 
\begin{align}
C_{-2,0} & =\left(\frac{x_{s}}{k_{1}}\right)^{2}\frac{(igA+2L-2M)}{(igA+N-2M+1)}
\end{align}
We could repeat the calculation for the other descendant
\begin{equation}
|\mathbf{y},\mathbf{w};-2,0^{*}\rangle^{b}:=\mathbb{Q}_{2}^{\dagger\,3}\mathbb{Q}_{2}^{\dagger\,4}|\mathbf{y},\mathbf{w}\rangle^{b}\qquad;\quad C_{-2,0^{*}}=\frac{A}{((2L-N-1)\frac{i}{g}-A)}
\end{equation}
Using similar calculations the normalization constants can be fixed.

\bibliographystyle{JHEP}
\bibliography{ref}

\end{document}